\documentclass[useAMS]{gGAF2e}

\begin{document}
\doi{10.1080/03091920xxxxxxxxx}
\issn{1029-0419} \issnp{0309-1929} 

\markboth{Vortical and Wave Modes... }{Sukhatme and Smith}

\title{Vortical and Wave Modes in 3D Rotating
Stratified Flows:
Random Large Scale Forcing }

\author{Jai Sukhatme$^1$\thanks{$^\ast$Corresponding
author. Email: sukhatme@math.wisc.edu\vspace{6pt}} and Leslie M. Smith$^{1,2}$ \\
1. Mathematics Department, University of Wisconsin-Madison, WI 53706  \\
2. Engineering Physics Department, University of Wisconsin-Madison, WI 53706}

\maketitle

\begin{abstract}

Utilizing an eigenfunction decomposition, we study the growth and spectra of energy in the 
vortical (geostrophic) and wave (ageostrophic) 
modes of a three-dimensional
(3D) rotating stratified fluid as a function of $\epsilon = f/N$, where $f$ is the Coriolis parameter 
and $N$ is the Brunt-Vaisala frequency. Throughout we employ a random large scale forcing in a unit aspect ratio domain 
and set these parameters such that the 
Froude and Rossby numbers are roughly comparable 
and much less than unity. Working in regimes characterized by moderate Burger numbers, i.e.\ 
$Bu = 1/\epsilon^2 < 1$ or $Bu \ge 1$, our results indicate 
profound change in the character of vortical and wave mode interactions with respect to $Bu = 1$.
Previous analytical work concerning the qualitatively different nature of these interactions has been
in limiting conditions of rotation or stratification domination (i.e.\ when $Bu \ll 1$ or $Bu \gg 1$ 
respectively).
As with the reference state of $\epsilon=1$, 
for $\epsilon < 1$ the wave mode energy saturates quite  
quickly and the ensuing forward cascade continues to act as an efficient means of dissipating ageostrophic energy. 
Further, these saturated spectra steepen as  $\epsilon$ decreases:  we see
a shift from $k^{-1}$ to $k^{-5/3}$ scaling for $k_f < k < k_d$ (where $k_f$ and $k_d$ are
the forcing and dissipation scales, respectively).
On the other hand, when $\epsilon > 1$ 
the wave mode energy never saturates and comes to dominate the total energy in the system. 
In fact, in a sense the wave modes behave in an asymmetric manner about $\epsilon = 1$.
With regard to the vortical modes, 
for $\epsilon \le 1$, the signatures of 3D quasigeostrophy are clearly evident. 
Specifically, we see a $k^{-3}$ scaling for $k_f < k < k_d$
and, in accord with an inverse transfer of energy, the vortical mode energy never 
saturates but rather increases for all $k < k_f$. 
In contrast, for $\epsilon > 1$ and increasing, the vortical modes contain a progressively
smaller fraction of the total energy indicating that the 3D quasigeostrophic subsystem
plays an energetically smaller role in the overall dynamics. Combining the vortical and 
wave modes, 
the total energy for $k>k_f$ and $\epsilon \le 1$ shows a transition 
as $k$ increases wherein the vortical modes contain a large portion of the energy
at large scales, while the wave modes dominate at smaller scales.
There is no such transition when $\epsilon > 1$ and the wave modes dominate the 
total energy for all $k > k_f$.

\end{abstract}

\section{Introduction}

In terms of the Froude ($Fr$) and Rossby ($Ro$) numbers,  
geophysical phenomena occur in an environment characterized by strong rotation and stratification, i.e.\ these numbers are
$O(10^{-1})$ or 
smaller (Gill 1982).   In the present work, we consider a forced scenario 
where $f, N$ (the Coriolis parameter
and Brunt-Vaisala frequency) are comparable, and thus define these parameters by
$Fr=(\epsilon_f k_f^2)^{1/3}/N$ and $Ro=(\epsilon_f k_f^2)^{1/3}/f$ where $\epsilon_f, k_f$ are the energy input rate and forcing wavenumber, respectively.  
The influence of rotation and stratification on fluid flows has been examined in detail, and 
recent reviews include Riley and Lelong (2000) for stratification, Cambon et al.\ (2004)
for rotation and
Cambon (2001) for both stratification and rotation. \\

For purely rotating flows ($\vec{\Omega}=\Omega \hat{z}$), the system supports inertial waves obeying the
dispersion relation $\sigma(\vec{k})= \pm 2\Omega k_z/k$.
When forced at small scales and for $Ro < O(1)$, 
the principal effect of rotation is seen in the two-dimensionalization
of the flow:  simulations indicate
the spontaneous formation of columnar structures from initially isotropic conditions 
(Smith \& Waleffe 1999, see Bartello, Metais \& Lesieur (1994) for analogous behavior in the decaying problem).
In effect, there is a transfer of energy to
the slow wave modes ($k_z=0 \Rightarrow \sigma(\vec{k}) = 0$) and this transfer has an inverse character.
Given that exactly resonant interactions cannot transfer energy to slow wave modes (Greenspan 1969), 
near-resonances play an important role in this process (Smith \& Lee 2005). 
Analytically, for $Ro \rightarrow 0$, the limiting dynamics involving purely resonant interactions 
contain three parts (Babin, Mahalov \& Nicolaenko 1999). The first part, consisting of zero frequency 
interactions, yields the two-dimensional (2D) Navier-Stokes equations. The second part, involving 
interactions of the fast-zero-fast type, results in the passive driving of the vertical velocity 
by means of the aforementioned 2D flow (Babin, Mahalov \& Nicolaenko 1996, Chen et al.\ 2005). 
Finally, including the third category of interactions involving 
three fast wave modes results in the so-called 2$\frac{1}{2}$ dimensional equations 
(Babin, Mahalov \& Nicolaenko 1999).
\footnote{Quite interestingly, a family of dynamically active dispersive scalars (which includes the 
familiar barotropic beta plane equations as a member) is seen to have a similar structure of resonant 
interactions (Sukhatme \& Smith 2007a). Indeed, there the analog of the present transfer to 
columnar structures is
the spontaneous generation of zonal flows.\\}
When forced at large scales (for $Ro < O(1)$), 
in agreement with the analytical picture of so-called "phase turbulence" 
for fast wave
modes with $\sigma \neq 0$ (Babin, Mahalov \& 
Nicolaenko 1996),  
the transfer of energy to
smaller scales is seen to be inefficient (Cambon et al.\ 1997, see also Bellet et al.\ 2006 for 
statistical closures of these wave-turbulence phenomena).
Also, for rapid rotation, along with an accumulation of energy near the forcing scale  
there is phenomenological (Zhou 1995) and numerical support 
for a $k^{-2}$ scaling in an intermediate range of scales (Yeung \& Zhou 1998, Bellet et. al.\ 2006). \\

For uniform vertical stratification, in addition to gravity waves satisfying
$\sigma(\vec{k})= \pm N k_h/k$ (where $k_h^2 = k_x^2 + k_y^2$) the equations possess a distinct zero frequency mode, the
so-called vortical mode (Lelong \& Riley 1991). 
With forcing at small scales and for $Fr < O(1)$, the principal effect of stratification is seen in the
layering of fields:
numerical simulations indicate the spontaneous emergence of layers perpendicular to
the ambient stratification (Metais \& Herring 1989, Smith \& Waleffe 2002; see Godeferd \& Cambon 1994, 
Riley \& De Bruyn Kops 2003 for
the 3D decaying case; see
Smith 2001 and Sukhatme \& Smith 2007b for, respectively, the forced 2D case and the decaying 2D case). 
As for pure rotation, there is an
an inverse transfer of energy to the slow wave modes ($k_h=0 \Rightarrow$ decoupled layers) (Smith \& Waleffe 2002).
In the limit $Fr\rightarrow 0$, the presence of a reduced system
consisting of diffusively coupled 2D layers is anticipated by a scaling
analysis of the original equations (Riley, Metcalfe and Weissman 1981, Lilly 1983, Majda \& Grote 1997).
With forcing at large scales and $Fr < O(1)$,  there are significant 
differences for the scaling of the energy spectra depending on the nature and dimensionality of the force
(Waite \& Bartello 2004, Lindborg 2006).
However, 
in accord with 
the analytical picture of "unfrozen" wave mode
cascades (Babin, Mahalov \& Nicolaenko 1997),
simulations employing these vastly differing forcing mechanisms agree upon the
robust nature of energy transfer
to small scales. \\ 

In the presence of both rotation and stratification, the focus of the present work, an appropriate governing framework is 
provided by the 
3D rotating Boussinesq equations. 
Note that the aforementioned results regarding rotating flows hold when $N=0$, and that 
it is not immediately clear how the limit of large $\epsilon=f/N$ 
converges to a purely rotating flow.
In essence the role of the extra vortical mode that exists whenever $N \neq 0$ needs careful consideration.\footnote{From recent discussions with Prof. A.J. Majda, the limit of $Ro \rightarrow 0$ while holding 
$Fr \sim O(1)$ appears to be described by a reduced 
system involving barotropic dynamics coupled with oscillatory density and vertical velocity fields. This 
is presently being examined in greater detail. \\}
On the other hand, the limit of small $Fr$ with fixed $Ro$ has
been rigorously analyzed and shown to obey a splitting between the
vortical and fast wave modes wherein both the vortical and slow wave modes evolve as 
vertically sheared horizontal flows (VSHF) (Embid \& Majda 1998).
Setting $f=0$ in this limiting case results in the previously mentioned small $Fr$ reduced system 
of diffusively coupled 2D layers for purely stratified
flows. \\

In the linear and non-dissipative limits, 
the 3D rotating Boussinesq equations support wave-like motions that obey the dispersion relation 
(see Majda 2003 for a recent overview)

\begin{equation}
\sigma_{\pm} (\vec{k}) = \pm  \frac{N ( k_h^2 + \epsilon^2 k_z^2)^{1/2}}{k}~;~ \sigma_0(\vec{k})=0
\label{1}
\end{equation}
where $\epsilon=f/N=Fr/Ro$. 
Here modes corresponding to $\sigma_{\pm}$ are called wave (or ageostrophic) modes
while the mode corresponding to $\sigma_0$ is referred to as 
the vortical (or geostrophic) mode (Bartello 1995, Smith \& Waleffe 2002).
In a unit aspect ratio domain, the Burger number is defined as $Bu=(Ro/Fr)^2$.
In the presence of strong stratification and rotation a regime relevant to 
atmosphere-ocean dynamics is
$Fr, Ro \ll 1$ with $Bu \sim 1$ (Gill 1982). A rigorous analysis of the 
limit $Fr, Ro \rightarrow 0$, $Bu \sim 1$ 
indicates a splitting between the vortical and wave modes of the 3D rotating Boussinesq system wherein
the vortical modes evolve as an independent 3D quasigeostrophic (QG)
system while the wave modes evolve via (catalytic) wave-vortical-wave and wave-wave-wave interactions 
(Babin et al.\ 1997, Embid \& Majda 1998). Bartello (1995) performed a detailed study
of geostrophic adjustment for $f \approx N$ with $\sigma_{\pm} \approx \pm N$, implying a system with
vortical and nearly constant frequency wave modes. Examining the decay of a large scale initial condition, 
he noted the aforementioned decoupling between the vortical and wave modes. Further, apart from the 
3D QG dominance of the vortical modes (accompanied by a $k^{-3}$ scaling) it was seen numerically
that catalytic wave-vortical-wave interactions dominated 
the wave mode dynamics (accompanied by a $k^{-1}$ scaling) resulting in a rapid transfer of energy to small scales. 
In this case, the catalytic interactions are responsible for the dissipation of ageostrophic energy and 
the emergence of a geostrophically adjusted state 
(Bartello 1995). In a later study, Waite \& Bartello (2006) found a gradual transition from 
stratified to quasigeostrophic turbulence as the Rossby number was decreased for a fixed small Froude 
number. Employing a small scale random forcing for $1/2 \le \epsilon \le 2$, evidence for 
the dominance of vortical modes by QG-like dynamics was also presented in Smith \& Waleffe (2002),
 who detected the presence of a 
$k^{-5/3}$ upscale energy cascade amongst the 3D vortical modes (see also Metais et al.\ 1996). \\

As $\epsilon$ gradually deviates from unity, the wave mode frequencies are non-zero and no longer constant, 
i.e.\  $f \le |\sigma_{\pm}| \le N$ (and vice versa if $N < f$). 
As is known (and will be elaborated on in the next section), the dependence of wave mode frequencies on wavenumber
places restrictions on the existing vortical-wave mode interactions while
allowing for additional interactions not present when $f=N $. 
As part of an extensive body of work, Babin et al.\ and Embid \& Majda have examined a variety 
of asymptotic limits for both the 
purely rotating and the rotating stratified Boussinesq equations; this work is
reviewed in Babin, Mahalov and Nicolaenko (2002) and Majda (2003) respectively. 
In particular, rigorous analysis with respect to the $Bu \ll 1$ and $Bu \gg 1$ limits (note that 
the relevant Froude and/or Rossby numbers are always assumed to be small) indicates a profound 
difference in the behavior of the 
vortical and wave modes in these 
rotation vs. stratification dominated cases, respectively. 
Indeed, based on these limiting scenarios, our aim is to examine if the difference in the behavior of the 
modes begins to gradually manifest itself as one 
moves in the direction $\epsilon > 1$ or $\epsilon < 1$. Of course, as $Bu = 1/\epsilon^2$, a moderate
deviation from unity in $\epsilon$ is much more substantial in terms of $Bu$. 
As suggested by the so-called "regime diagram" in the review by 
Babin et al.\ (2002), 
even though rotation domination is analytically justifiable only in the limit of small $Bu$, it
is interesting to investigate how quickly rotation dominates as 
the Burger number decreases from unity
(and vice versa for the stratification dominated regime). In fact, these intermediate regimes in $Bu$
are precisely those that are of geophysical interest.  Further, intermediate $Bu$ regimes 
are not presently accessible to theoretical analysis, hence
motivating a numerical investigation. \\

Utilizing a random, isotropic, large scale forcing that projects equally on all the modes 
and keeping both $Fr,Ro \ll O(1)$, 
our objective in the present work is to test if we can detect a
difference in the behavior of the various modes 
with respect to $\epsilon =1$, with $\epsilon=1$ as a well understood
reference state (Bartello 1995). The reasoning behind employing an isotropic random force is to avoid
biasing any specific mode, i.e.\ the system is allowed to choose the dominant response.
Specifically, numerical simulations are used to investigate how
varying $\epsilon$ (such that we go from $Bu = 1/25$ to $Bu = 25$) 
affects the scaling of energy spectra, the growth of energy and the overall role played by 
the vortical and wave modes of the 3D rotating Boussinesq system. 
Further, this lack of severe anisotropy 
enables us to confidently
use a numerical grid with uniform spacing in all directions. Indeed, if one proceeds to extremes in
$\epsilon$ it is quite important to consider skewed grids that are able to properly 
resolve the dynamics (see for example Lindborg 2006 for the strongly stratified case).
In the next section we introduce
the basic equations and their limiting form for $Ro, Fr \rightarrow 0$ as 
an eigenfunction decomposition.  From the latter, one anticipates
a change in the dynamics with $\epsilon$. In Section III we present 
a systematic numerical examination as $\epsilon$ deviates from unity. 
For a fair comparison, all simulations are carried out 
with a fixed input energy rate and varying $f, N $ so as to keep both 
$Ro, Fr \ll O(1)$. Finally in Section IV we collect and discuss the results. 

\section{The basic equations and their spectral decomposition } 

The rotating Boussinesq equations in a 3D periodic setting read 

\begin{eqnarray}
\frac{D \vec{u}}{Dt} + f \hat{z} \times \vec{u} = - \nabla \phi - N \theta \hat{z} + \nu \nabla^2 \vec{u} \nonumber \\
\frac{D \theta }{Dt} - N w = \kappa \nabla^2 \theta 
\label{2}
\end{eqnarray}
where $\vec{u}=(u,v,w)$ is the 3D velocity field, $\phi=p/\rho_0$ is a scaled pressure and $f$ is the Coriolis parameter 
with rotation assumed to be aligned with the $z$-axis.  Equations (\ref{2}) result from 
considering periodic perturbations to a state of hydrostatic balance wherein the density profile satisfies
$\rho = \rho_0 - b z + \rho'$ with $|\rho'|, |bz| \ll \rho_o$. 
Further, we have set $\rho'= (N \rho_o/g) \theta$ where $N=(gb/\rho_0)^{1/2}$.
Finally, $\nu$ and $\kappa$ are the viscosity and diffusivity respectively. 
The solution to the general linear initial value problem can be represented as (see Embid \& Majda 1996, 1998
for a detailed spectral analysis of the linear operator arising from (\ref{2})) 

\begin{equation}
\hat{\Psi}(\vec{x},t) = \sum_{\vec{k}} \sum_{\beta=0,\pm } a_{\beta}(\vec{k},t) \phi_{\beta}(\vec{k}) \exp{[i \vec{k} \cdot \vec{x} ]} 
\label{3}
\end{equation}
where $\hat{\Psi}=(\hat{u} ~\hat{v} ~\hat{w} ~\hat{\theta})^{T}$, $a_0, a_{\pm}$ are determined by the 
initial/forcing conditions and $\phi_0,\phi_{\pm}$ are the mutually orthogonal eigenfunctions corresponding to $\sigma_0,\sigma_{\pm}$ 
(the forms of the eigenfunctions can be found in Embid \& Majda 1999 and Smith \& Waleffe 2002, see also
Bartello 1995 for an alternate
construction). 
As the basis formed by these eigenfunctions is complete, we use it to expand the solution to the nonlinear problem in (\ref{2})
(Embid \& Majda 1998, Smith \& Waleffe 2002).
Symbolically the nonlinear coupling yields

\begin{equation}
\frac{d a_{\alpha}(\vec{k})}{dt} + i \sigma_{\alpha}(\vec{k}) a_{\alpha}(\vec{k}) = 
\sum_{\triangle} \sum_{\beta,\gamma} C_{\vec{k}\vec{p}\vec{q}}^{\alpha \beta \gamma}~ {a}_{\beta}(\vec{p}) {a}_{\gamma}(\vec{q})
\label{4}
\end{equation}
where $\triangle$ represents a sum over $\vec{p},\vec{q}$ such that $\vec{k}=\vec{p}+\vec{q}$. The indices $\alpha,\beta,\gamma$ run over 
$0,+,-$
and $C_{\vec{k}\vec{p}\vec{q}}^{\alpha \beta \gamma}$ is the interaction 
co-efficient. 
Before proceeding we introduce some notation: 
triplets $(\cdot, \cdot, \cdot)$ refer to a particular type of triad interaction, where
$z$ refers to the vortical mode and $f$ denotes the wave mode (not to be confused
with the vertical coordinate and the Coriolis parameter, respectively).  For example, 
$(z,f,f)$ represents the evolution of a vortical mode via the interaction
of two wave modes (i.e.\ the first entry evolves via an interaction of the latter two entries in the triplet). 

\subsection{Resonant and near-resonant interactions}

Setting $a_{\alpha} = b_{\alpha} ~\exp{(-i \sigma_{\alpha}(\vec{k}) t)}$ where $\alpha=\{0,+,-\}$ in (\ref{4}), the so-called limiting dynamics for
$Fr,Ro \rightarrow 0$ --- due to the cancellations amongst the oscillatory phase factors --- are given 
exclusively by 
purely resonant interactions,

\begin{equation}
\frac{d b_{\alpha}(\vec{k})}{dt} =
\sum_{\triangle_R} \sum_{\beta,\gamma} C_{\vec{k}\vec{p}\vec{q}}^{\alpha \beta \gamma}~ {b}_{\beta}(\vec{p}) {b}_{\gamma}(\vec{q})
\label{l2}
\end{equation}
where $\triangle_R$ restricts the sum to triads with zero sum of wave frequencies as
well as wavevectors 
(the reader is referred to Embid \& Majda 1996, 1998 for a formal derivation of (\ref{l2}) via a two timescale analysis, 
see also Anile et al.\ (1993) for an overview of such techniques in the context of 
general nonlinear dispersive equations).
The vortical modes evolve as

\begin{equation}
\frac{d b_0(\vec{k},t)}{dt} = \sum_{\triangle} ~C_{\vec{k}\vec{p}\vec{q}}^{0 0 0} ~{b}_{0}(\vec{p}) {b}_{0}(\vec{q})
+ \{C_{\vec{k}\vec{p}\vec{q}}^{0 \pm \mp} ~{b}_{\pm}(\vec{p}) {b}_{\mp}(\vec{q})\}^
{[\sigma_{\pm}(\vec{p})+\sigma_{\mp}(\vec{q}) = 0]} 
\label{7}
\end{equation}
while the 
wave mode evolution is given by 

\begin{eqnarray}
\frac{d b_{+}(\vec{k})}{dt} = \sum_{\triangle} 
\{C_{\vec{k}\vec{p}\vec{q}}^{+ 0 +} ~{b}_{0}(\vec{p}) {b}_{+}(\vec{q}) \}^{[\sigma_{+}(\vec{q})-\sigma_{+}(\vec{k}) = 0]} +
\{C_{\vec{k}\vec{p}\vec{q}}^{+ + 0} ~{b}_{+}(\vec{p}) {b}_{0}(\vec{q}) \}^{[\sigma_{+}(\vec{p})-\sigma_{+}(\vec{k}) = 0]} + \nonumber \\
\{C_{\vec{k}\vec{p}\vec{q}}^{+ + \pm} ~{b}_{+}(\vec{p}) {b}_{\pm}(\vec{q}) \}^{[\sigma_{+}(\vec{p})+\sigma_{\pm}(\vec{q})-\sigma_{+}(\vec{k}) = 0]} +
\{C_{\vec{k}\vec{p}\vec{q}}^{+ - \pm} ~{b}_{-}(\vec{p}) {b}_{\pm}(\vec{q}) \}^{[\sigma_{-}(\vec{p})+\sigma_{\pm}(\vec{q})-\sigma_{+}(\vec{k}) = 0]}
\label{8}
\end{eqnarray}
and a equation analogous to (\ref{8}) governs $b_-(\vec{k})$. 
As we are interested in moderate values of $\epsilon$ with nonzero
wave frequencies $\sigma_{\pm} \in [f,N]$, the 
$(z,z,f)$ and $(f,z,z)$ interactions do not appear in (\ref{7}) and (\ref{8}) respectively
as they can never be 
resonant. \\

It is important to note that (\ref{7}), (\ref{8}) are limiting equations and 
do not account for all of the energy transfer 
among the different modes in the finite $Ro, Fr$ system. In particular, for 
$Fr, Ro$ small but non-zero,
near-resonances (or quasi-resonances) play a significant role in the transfer of energy
in addition to exact resonances. 
\footnote{In periodic domain, the importance of near-resonances 
is further highlighted by the
fact that it may be impossible for a triad to satisfy the dispersion relation when the admissible wavenumbers are 
restricted to $\mathbb{Z}$ (Kartashova 1994) --- see Connaughton et al.\ (2001) and Lvov et al.\ (2006)
for examples.\\}
For example, consider a discrete initial value problem. For near resonances with
wave frequencies summing to $\delta \ll 1$, and 
denoting $\min(Ro,Fr)$ by $\gamma \ll 1$, 
then the phase factors implicit in (\ref{4}) oscillate with periods $\sim \delta/\gamma$.  
These near-resonances will contribute until $t \sim \gamma/\delta$ and
some routes of energy transfer
that are precluded under resonance
can potentially be adopted via these near-resonant interactions, hence altering 
predictions from purely resonant considerations (an example of which is provided by Watson \& Buchsbaum 1996).
\footnote{Of course, there is also the possibility of successive resonant triads leading to quartet interactions (Newell 1969). \\}
In a periodic domain with small-scale forcing, Smith \& Lee (2005) have
demonstrated that near-resonant triads are responsible for 
the spontaneous generation of asymmetric large scale columnar structures 
from small-scale fluctuations in purely rotating 3D flows.
In fact, by broadening the resonance condition 
in the kinetic equation, 
near-resonant interactions have recently been shown to have important consequences in
statistical theories for the evolution of random wave fields (Janssen 2003, Annenkov \& Shrira 2006). 
The present work, as with the foundational references, assumes an inherent 
heirarchy wherein much of the dynamics can be understood via exact and 
near-resonant interactions. 
Of course, in reality there are contributions from interactions that are far
from resonance.
Section 2.2 discusses exact resonances in the
case $\epsilon=1$ (for which there are no near-resonances), and Section 2.3 
discusses both exact and near-resonances for $\epsilon \neq 1$.  In Section 3, the simulation
results for the full dynamical equations (4) indicate that exact and near-resonances
are indeed helpful to explain the evolution of energies and spectra associated with 
the dynamics for the parameter regimes
considered herein.

\subsection{$\epsilon = 1$ : Vortical and constant frequency wave modes}

It is known that resonant interactions involving two wave and one vortical mode are catalytic
in the sense that 
the vortical mode acts as a mediator of energy exchange between the wave modes but does not change itself (Bartello 1995,
Embid \& Majda 1998).
Keeping this mind, for the special case of
for $\epsilon =1$ (i.e.\ $\sigma_{\pm}=\pm N$),
(\ref{7}) and (\ref{8}) simplify to 

\begin{eqnarray}
\frac{d b_0(\vec{k},t)}{dt} = \sum_{\triangle} ~C_{\vec{k}\vec{p}\vec{q}}^{0 0 0} ~{b}_{0}(\vec{p}) {b}_{0}(\vec{q}) \\
\frac{d b_{+}(\vec{k},t)}{dt} = \sum_{\triangle} \{ ~C_{\vec{k}\vec{p}\vec{q}}^{+ 0 +} ~{b}_{0}(\vec{p}) {b}_{+}(\vec{q})
+ ~C_{\vec{k}\vec{p}\vec{q}}^{+ + 0} ~{b}_{+}(\vec{p}) {b}_{0}(\vec{q}) \} 
\label{9}
\end{eqnarray}
and we obtain Bartello's equations (see Eqn 19 in Bartello 1995).
Note that, because there is no possibility of near-resonances when $\epsilon =1$, 
the exchange of energy 
between the vortical and wave modes is likely to be a minimum, making the
case $\epsilon=1$ quite possibly optimal in 
the sense of dynamical splitting between vortical and wave modes even at non-zero $Fr,Ro$. \\

In (\ref{9}), the
vortical modes evolve as an independent 3D QG system,
 while the wave modes are advected via this QG flow 
 (see Smith \& Waleffe 2002 for an explicit calculation
showing that the $(z,z,z)$ interactions lead to the 3D QG 
potential vorticity (PV) equation). 
Given that the vortical modes follow the 3D QG PV equation, we expect a
large fraction of the energy entering these
modes to flow upscale (Charney 1971).
Further, the equations governing the wave modes are the same as those for a passive scalar and
we have a situation wherein
these uncoupled wave modes are advected by a predominantly large scale
QG flow. Further, this situation where the bulk of the energy of the advecting flow is at large scales 
qualitatively corresponds to the
Batchelor regime for passive scalars (Batchelor 1959, Kraichnan 1974). 
This picture was verified by Bartello's (1995) decaying simulations.
Starting from a (fairly) random large scale initial condition (characterized by a scale $l$) and setting 
$f \approx N$ in (\ref{2}), it was seen that 
the energy in the vortical (geostrophic) modes moved upscale. 
For scales smaller than $l$, there emerged a transient $k^{-3}$ 
potential enstrophy transfer spectrum. 
Simultaneously, the energy in the wave (ageostrophic) modes moved downscale to be 
dissipated and resulted in a
transient $k^{-1}$ (passive scalar like) spectrum. 

\subsection{$\epsilon \neq 1$ : Vortical and dispersive wave modes}

As the wave mode frequencies begin to depend on wavenumber, not only do other classes of resonant interactions 
become permissible (see below) in the 
evolution of the wave and vortical modes, 
the aforementioned near-resonant interactions
also enter the picture.  
Indeed, when $Fr,Ro$ are small (though $\neq 0$) and unequal, 
the vortical and wave modes are expected to evolve by and large independently, but we do expect some transfer of energy 
from the vortical to wave modes and
vice versa due to the non-catalytic nature of the 
$(z,f,f)$ and $(f,z,f)$ near-resonances.
\footnote{As an aside, this exchange of energy between the different modes connects to the 
issue of the non-existence of an invariant slow manifold for the rotating Boussinesq system --- 
see for example Vanneste (2006), and 
the references therein, for a recent discussion of this subject.\\} \\

{\it Vortical modes: } It is expected that as long as both $Fr,Ro$ are small and comparable the vortical mode 
dynamics will be close to quasigeostrophy --- as mentioned, these are the essential ingredients required for the rigorous 
demonstration of 3D QG dominance of the vortical modes (Babin et al.\ 1997, Embid \& Majda 1998).
But note that in the $Fr \rightarrow 0$, fixed $Ro$ limit (i.e.\ stratification dominated regime) the vortical modes 
(along with the $k_h=0$ slow wave modes) 
evolve as a vertically sheared horizontal flow (Embid \& Majda 1998). 
In contrast for $Fr=\infty$, small $Ro$ (purely rotating regime),
the vortical mode in fact no longer exists. Of course these are extreme cases, but 
they do prompt an inquiry into the 
behavior of the vortical modes as $\epsilon$ deviates from unity. \\

{\it Wave modes: } 
As compared to the $f=N$ case, we now have the added possibility of $(f,f,f)$ resonant and 
near-resonant interactions.
However, exact $(f,f,f)$ resonances are difficult to achieve 
with discrete wavevectors given the algebraic complexity of simultaneously
satisfying the triad and dispersion relations.
In addition, they are only possible when $\epsilon < 1/2$ or
$\epsilon > 2$ (Smith \& Waleffe 2002). 
Further, for $Bu \sim O(1)$, the analysis of Babin et al.\ (2002) indicates that the set
of $(f,f,f)$ near resonances is very sparse. 
\footnote{Such estimates are required for investigating the issue of global regularity for the 3D rotating Boussinesq system, 
recently tackled by Babin et al.\ (2002). 
From a physical perspective, it is the forward transfer of energy 
that is likely to cause a breakdown in 
smoothness of the solutions and a control on these forward transfers is essential 
for any such investigation.\\} Hence for moderate values of $\epsilon$, it is unlikely 
that these additional resonant and near-resonant $(f,f,f)$ interactions would a cause significant 
deviation from the behavior seen for $\epsilon=1$ with constant wave mode frequencies.\\ 

With regard to the $(f,z,f)$ interactions,
as $\epsilon$ deviates from unity, 
these resonant and near-resonant interactions are restricted to wave modes that 
satisfy $|k_z/k_h| \approx |q_z/q_h|$.
Thus the $(f,z,f)$ exact and near-resonances for $\epsilon \ne 1$ are not as numerous
as catalytic $(f,z,f)$ exact resonances for the case $\epsilon=1$. 
In purely rotating flow, where the vortical mode no longer exists
and $z$ denotes a zero frequency wave mode, this restriction (which now becomes
$k_z/|k| \approx q_z/|q|$) confines the 
$(f,z,f)$ interactions only to re-distribute energy among similar scales (Babin et al.\ 2002).
The result is the so-called frozen cascade with an inefficient ageostrophic transfer
of energy from large to small scales. Indeed, the inefficiency of energy transfer to 
small scales has been repeatedly noted in numerical and statistical closure studies of purely 
rotating turbulence 
(Cambon et al.\ 1997, Yeung \& Zhou 1998 and Bellet et al.\ 2006). 
Even though the depleted transfer to small scales is seen for $\epsilon \gg 1$, this 
leaves us with the curious problem of how the change in the wave mode behavior arises as one 
eases away from $\epsilon =1$. 
In contrast, for a stratification dominated regime the
presence of the vortical mode ($z$) leads to the "unfreezing" 
of the ageostrophic forward transfer by the $(f,z,f)$ resonant and near-resonant interactions 
(Babin et al.\ 2002) --- 
in the special $f=N$ case this unfreezing manifests itself in catalytic $(f,z,f)$ resonances that,
as discussed, lead to the rapid dissipation of ageostrophic energy (Bartello 1995). \\

\section{ Numerical work } 

The numerical scheme to solve (\ref{2}) is a de-aliased pseudo-spectral method
implemented in a unit aspect ratio domain at a resolution of $256 \times 256 \times 256$. 
Time stepping is carried out by a third order Runge-Kutta method. 
We restrict the forcing to low wavenumbers (large scales) $3 \leq k \leq 4$. 
Further, the forcing is white in time and is chosen from a Gaussian
density function. The rate of input energy (denoted by $\epsilon_f$) in all cases is approximately 
the same, while the values of $f,N$ are changed so as to 
keep both $Ro,Fr \ll O(1)$ and achieve the desired $\epsilon$. 
Specifically, fixing $\epsilon_f$, 
we start with $(Fr,Ro)=(0.01,0.01)$ for $\epsilon =1$; 
for $\epsilon < 1$ we choose $(Fr,Ro)=(0.01,0.02),(0.01,0.03),(0.01,0.04), 
(0.01,0.05)$; vice versa for $\epsilon > 1$. \\

The only dissipation is small scale
hyperdiffusivity given by $(\nabla^2)^8$ (with $\nu=\kappa$).  Hyperdiffusivity allows for a range of large 
and intermediate scales unaffected by small- scale damping.    As one expects 
growth in the vortical mode energy for low wavenumbers $k<k_f$, the entire flow will not achieve a 
statistically steady
state without large scale damping. However, since the
the interaction of large scale damping with large scale wave modes is not well understood, we 
instead allow large scales to evolve freely.  We terminate all simulations when
the energy in wavenumbers $k<k_f$ is comparable to the energy in forced modes,
after which time finite size effects are expected to influence the dynamics.  

\subsection{Wave modes} 

\begin{figure}
\centerline{\epsfxsize=10cm \epsfysize=10cm \epsfbox{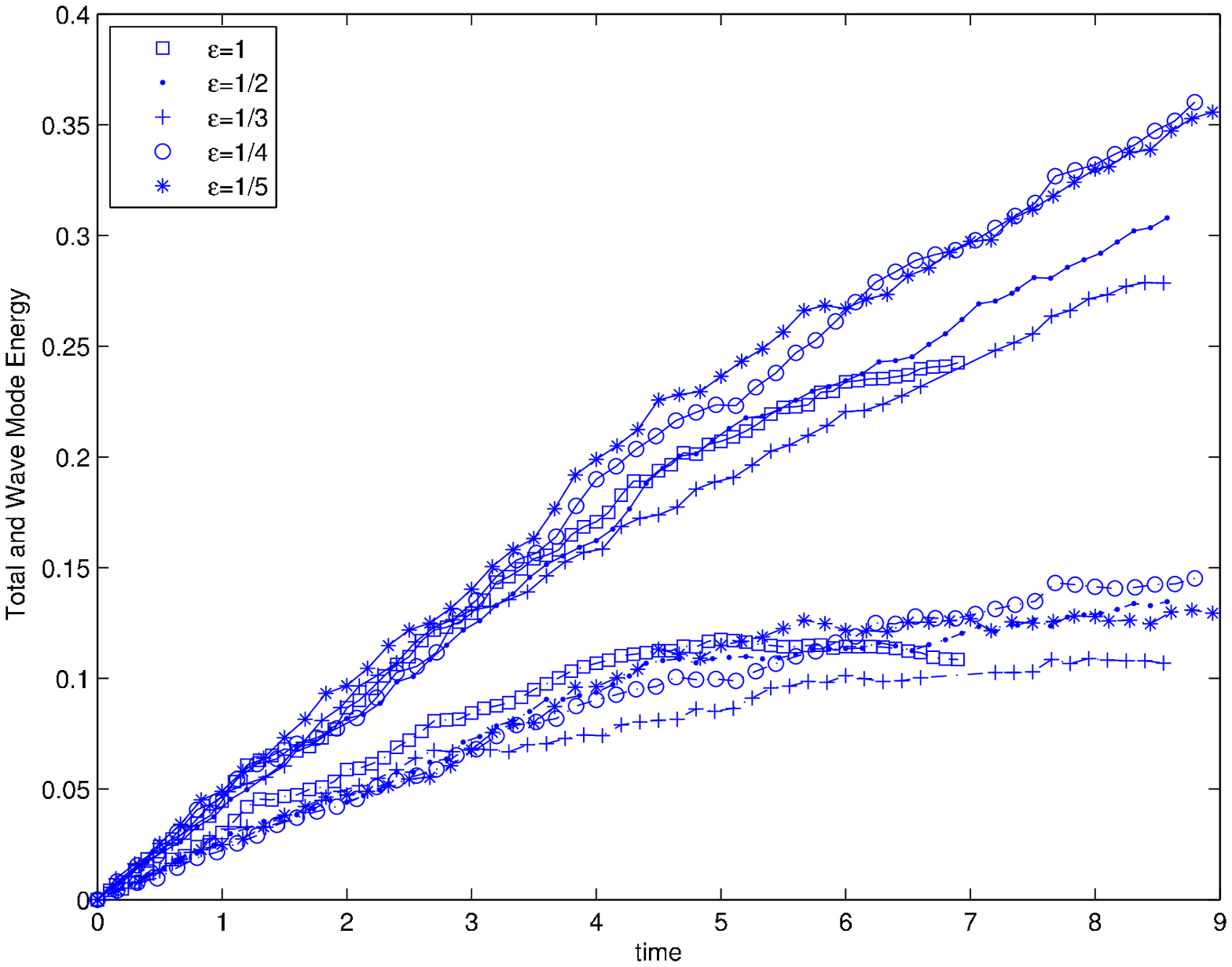}}
\centerline{\epsfxsize=10cm \epsfysize=10cm \epsfbox{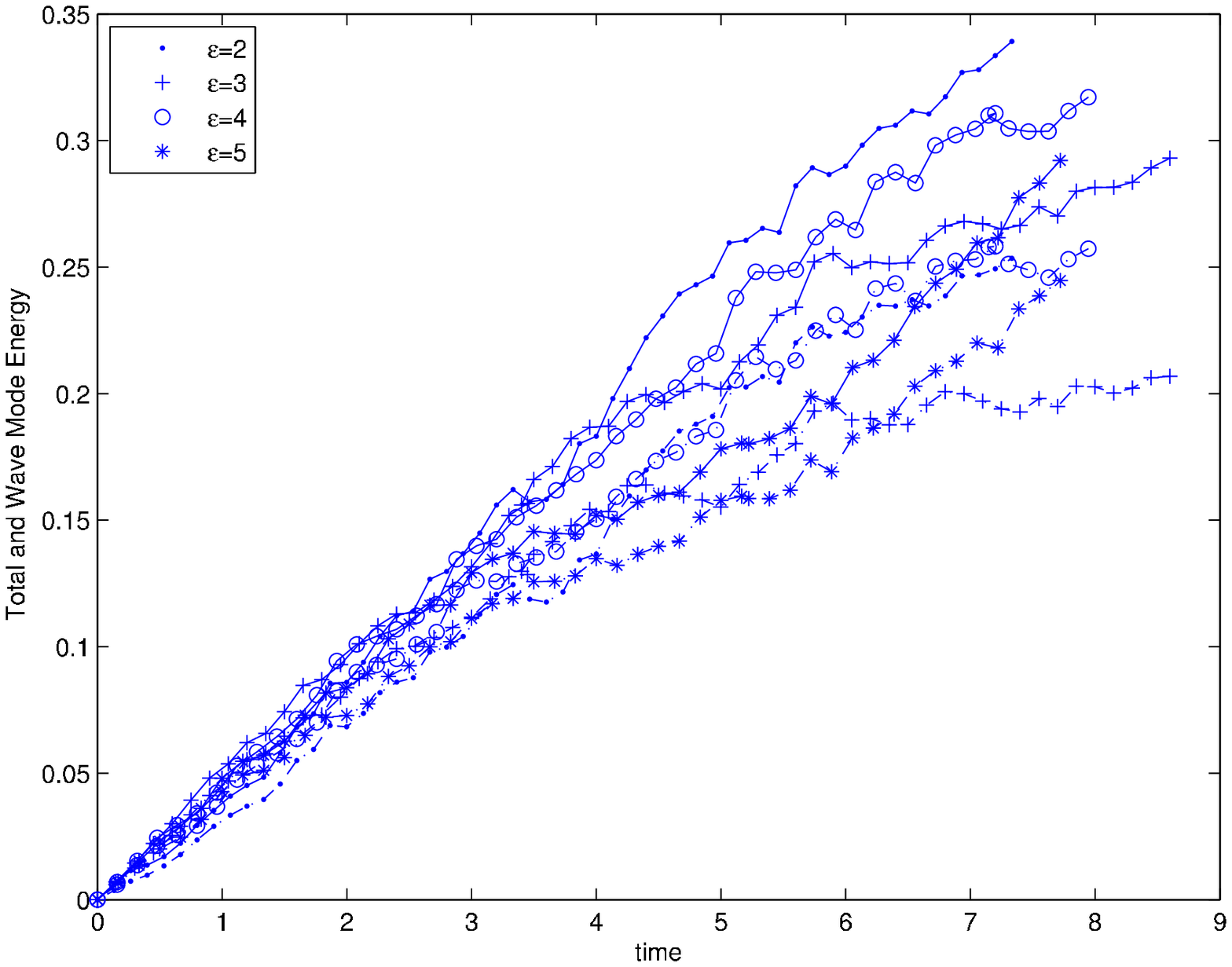}}
\caption{
Wave-mode and total energies as functions of time. Upper Panel:  $\epsilon \le 1$. The dash-dot lines with 
symbols are the wave-mode energies (lower bunch) while the solid lines with symbols (upper bunch) are the total energies. The saturation of wave-mode energy in all cases is quite evident. Second panel: $\epsilon > 1$ (same
notation). As is evident, the 
wave modes account for most of the energy in the system. }
\label{fig1}
\end{figure}

Fig. (\ref{fig1}) shows the wave-mode and total energies 
as functions of 
time. As is evident, for $\epsilon \le 1$ the wave mode energy saturates. In marked contrast, 
for $\epsilon > 1$, the wave modes soon come to dominate the energy in the system. 
Further, comparing the 
fraction of energy in the wave and vortical modes, Fig. (\ref{fig2}) indicates that the vortical mode dominance for
$\epsilon \le 1$ begins at an earlier time as stratification becomes stronger. 
On the other hand, for $\epsilon >1$,
dominance of the wave modes is more pronounced with increasing rotation.\\

Fig. (\ref{fig3}) shows the wave mode energy spectra for 
$\epsilon \le 1$ after saturation (see the caption for details). Closely 
following Bartello's observations in the decaying $f=N$ case (Bartello 1995), for $\epsilon =1$ the 
scaling attains a $k^{-1}$ form for $k_f < k < k_d$ consistent
with the passive Batchelor regime (Batchelor 1959, Kraichnan 1974). Interestingly,
as $\epsilon$ becomes smaller, the wave mode spectra steepen and approach a
$k^{-5/3}$ scaling by $\epsilon = 1/5$, i.e.\ for $(Fr,Ro)=(0.01,0.05)$.
For $\epsilon > 1$, even though the 
wave mode energy never saturates, the spectra appear to attain a near-invariant 
shape (first panel of Fig. (\ref{fig4})), i.e.\ they keep shifting upwards as the energy 
increases.  Note that, 
like pure rapid rotation (Yeung \& Zhou 1998), for larger values of $\epsilon$ there is some indication 
that a power law scaling may be emerging for
an intermediate range of scales.
Comparing the wave mode spectra for $\epsilon < 1$ and $\epsilon > 1$, it is noticeable that the peak at the 
forcing scale is much 
more pronounced when $\epsilon > 1$, i.e.\ rotation inhibits energy transfer out of the forcing scale. 
Irrespective of $\epsilon$, the energy at wavenumbers
$k< k_f$ does not grow significantly indicating the universal forward
nature of energy transfer amongst the wave modes. 

\subsection{Vortical modes} 

\begin{figure}
\centerline{\epsfxsize=10cm \epsfysize=10cm \epsfbox{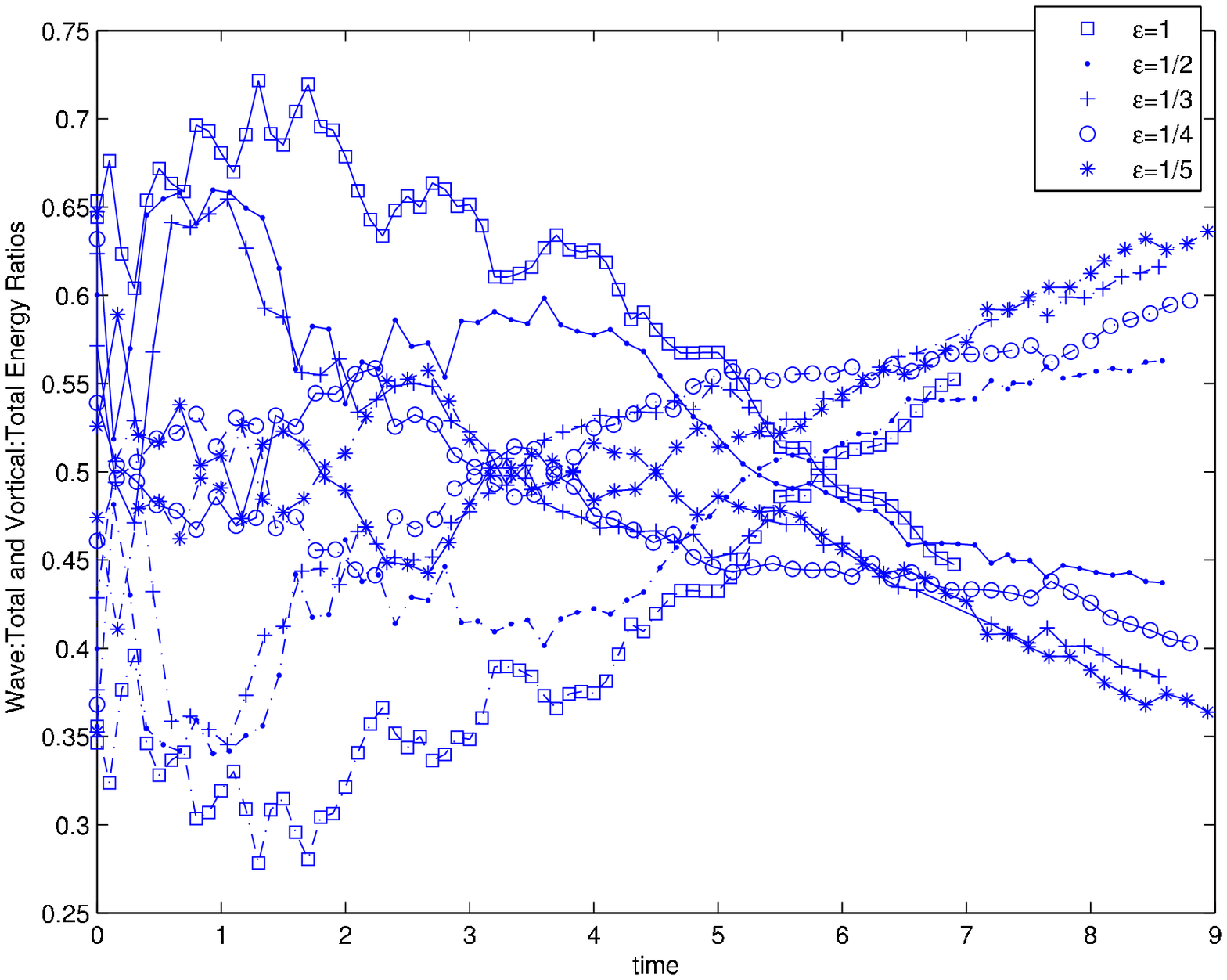}}
\centerline{\epsfxsize=10cm \epsfysize=10cm \epsfbox{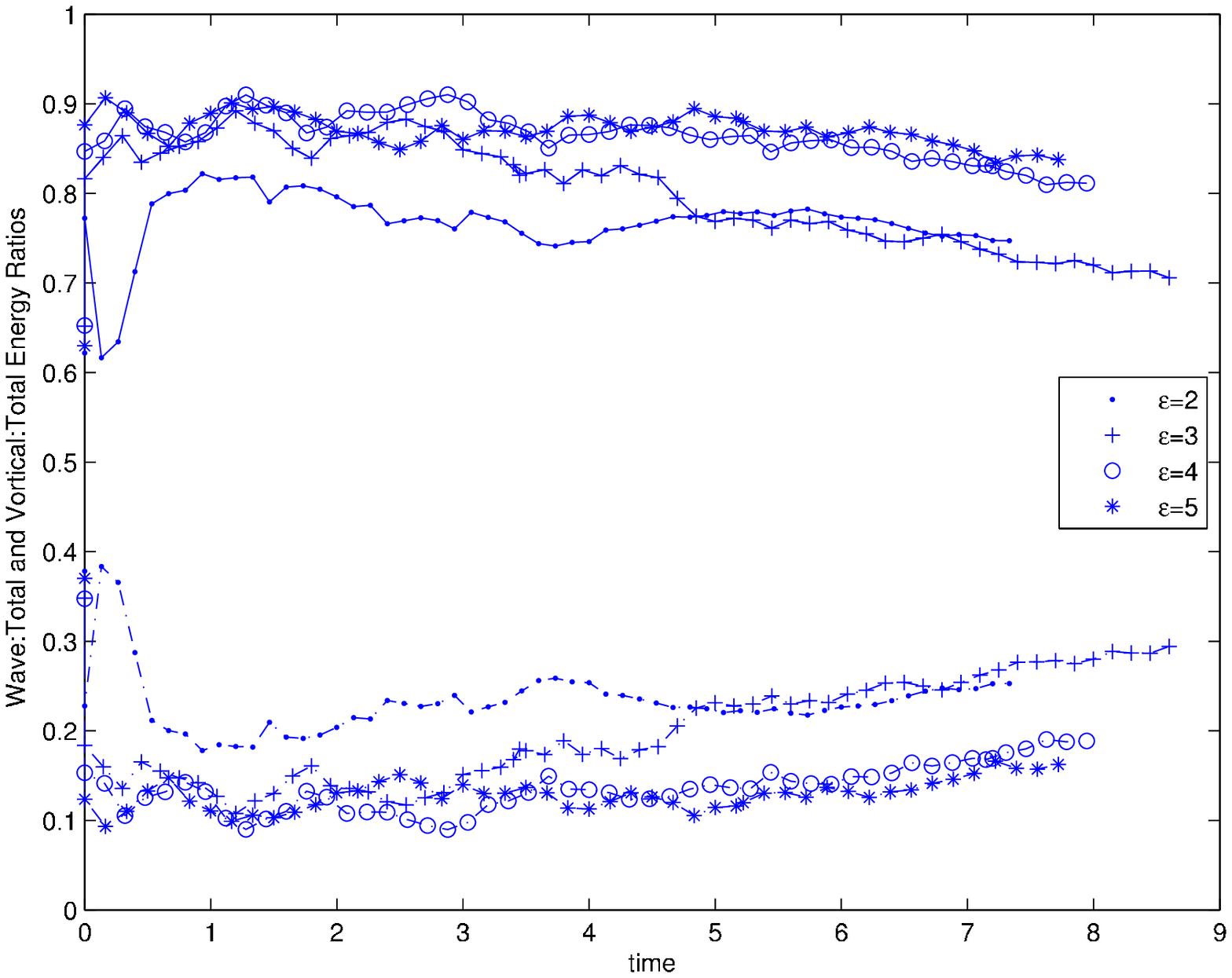}}
\caption{
Ratios of wave-mode energy to total energy 
(solid lines with symbols) and vortical-mode energy to total energy (dash-dot line with symbols). 
For $\epsilon \le 1$ (upper panel), initially the wave modes have a larger fraction of the total energy but
after a short time, as the wave modes saturate, 
the vortical modes proceed to contain a larger fraction of the total energy in the system. 
In contrast for $\epsilon > 1$ (lower panel), the wave modes dominate the energy budget and in fact, 
the fraction of energy in the vortical modes decreases with 
increasing $\epsilon$, i.e.\ as the rotation becomes stronger.}
\label{fig2}
\end{figure}

\begin{figure}
\centerline{\epsfxsize=10cm \epsfysize=10cm \epsfbox{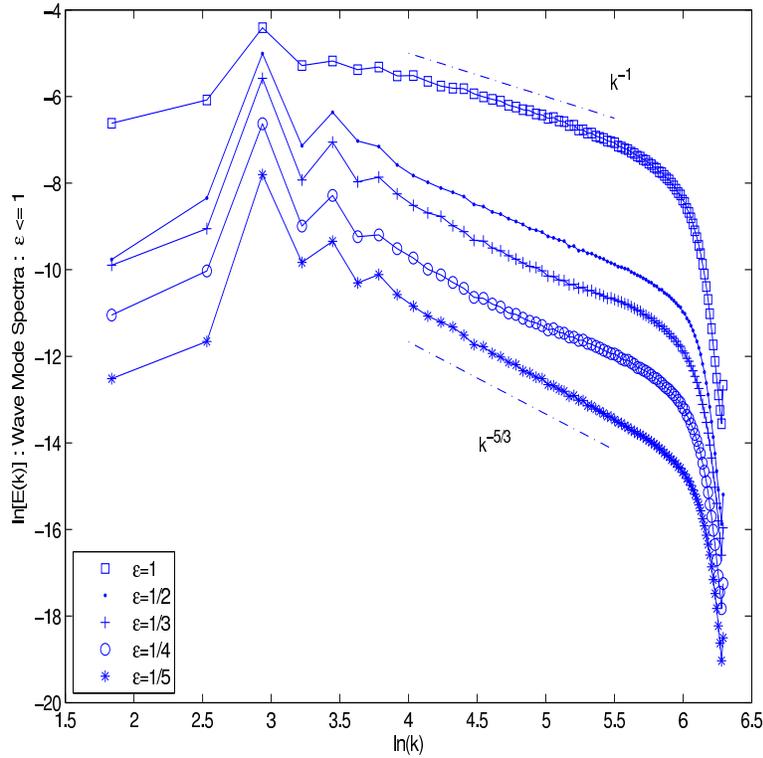}}
\caption{
Wave mode spectra with $\ln(k_f) \approx 3.2$ for $\epsilon \le 1$. Here,
$Fr=0.01, Ro=0.01 \rightarrow 0.05$. 
The spectra are plotted (vertically shifted for clarity) after the wave mode energy saturates, and are seen to 
change from a $k^{-1}$ to a $k^{-5/3}$ scaling for $k_f < k < k_d$ as $\epsilon$ decreases from
unity. At the start of each 
simulation the spectra are very steep (concentrated near the forcing scale) and become progressively
shallower.  
As the energy in the wave modes saturates, the spectra cease to evolve and attain the invariant shape 
plotted here. The simulations are halted when energy starts to accumulate in the large-scale
vortical modes so as to avoid finite size effects. }
\label{fig3}
\end{figure}

\begin{figure}
\centerline{\epsfxsize=10cm \epsfysize=10cm \epsfbox{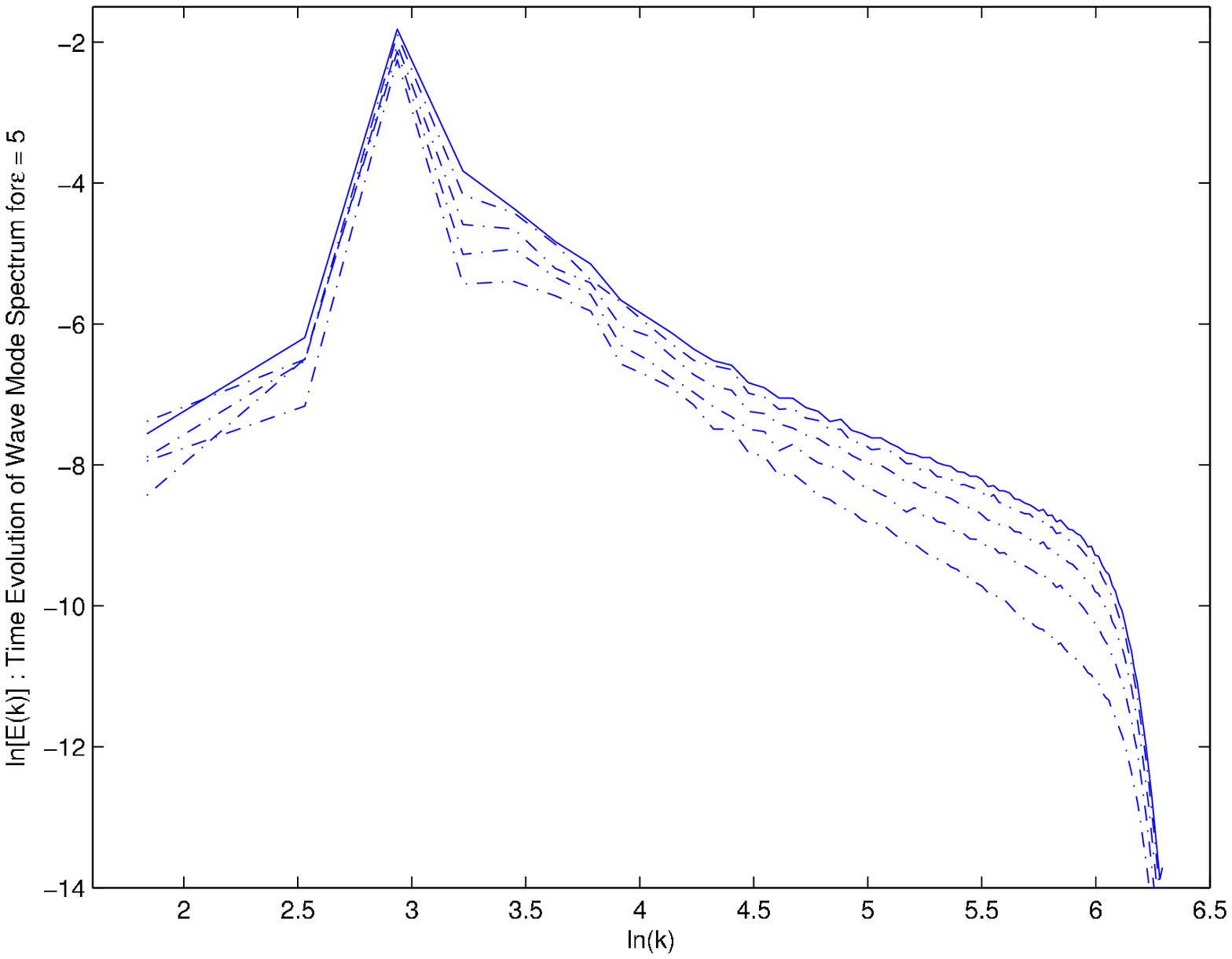}}
\centerline{\epsfxsize=10cm \epsfysize=10cm \epsfbox{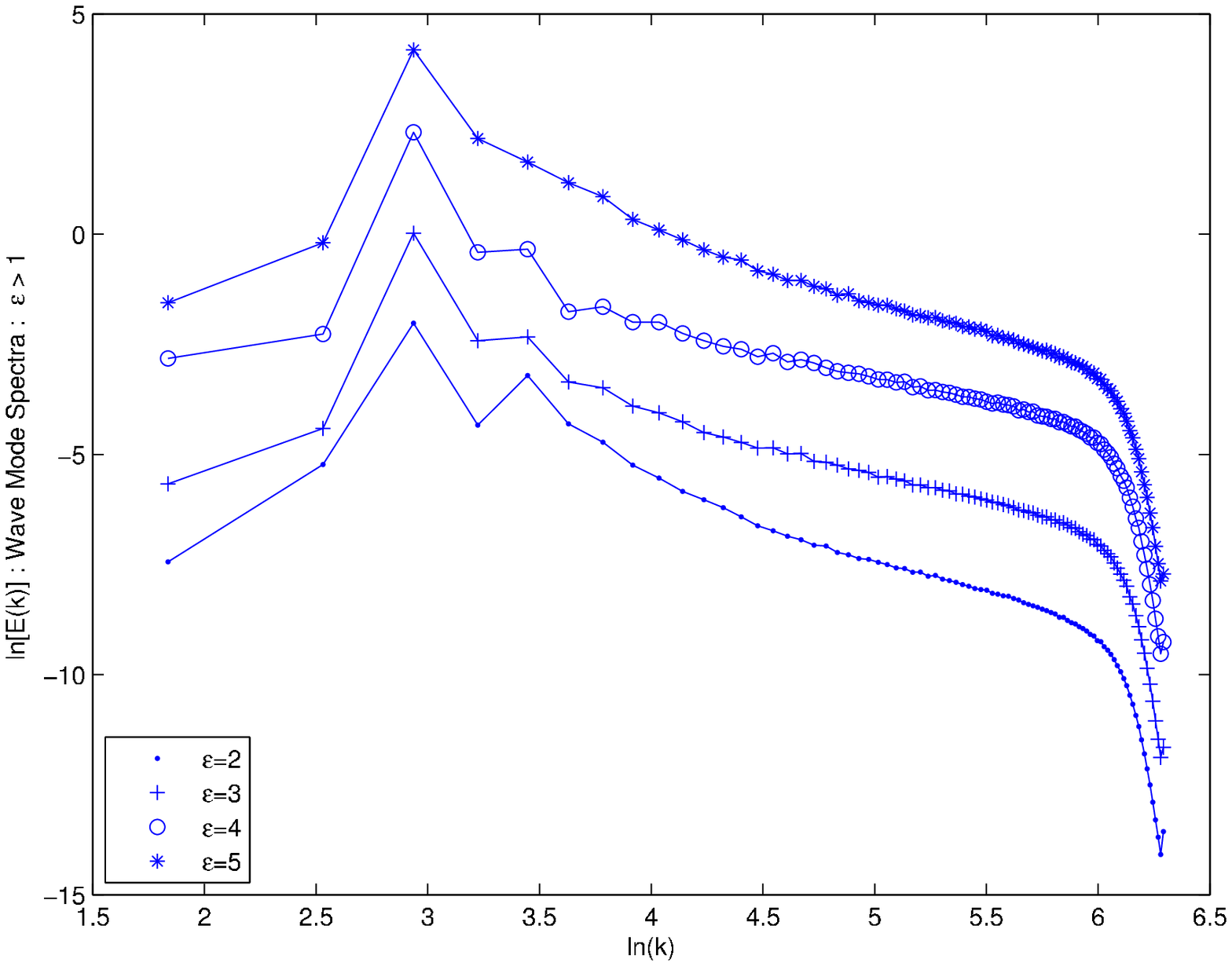}}
\caption{
Wave mode spectra with $\ln(k_f) \approx 3.2$ for $\epsilon > 1$. Here
$Ro=0.01, Fr=0.01 \rightarrow 0.05$. The upper panel shows the time evolution of the spectra for 
$\epsilon = 5$. Note that even though the energy keeps growing, the spectra appear to achieve an invariant 
shape. The lower panel shows the other $\epsilon > 1$ cases --- for clarity they have been vertically shifted.
Akin to purely rotating flows, for large $\epsilon$ there is some sign of a steeper than $5/3$ power law emerging at 
an intermediate range of scales.}
\label{fig4}
\end{figure}

\begin{figure}
\centering
\centerline{\epsfxsize=10cm \epsfysize=10cm \epsfbox{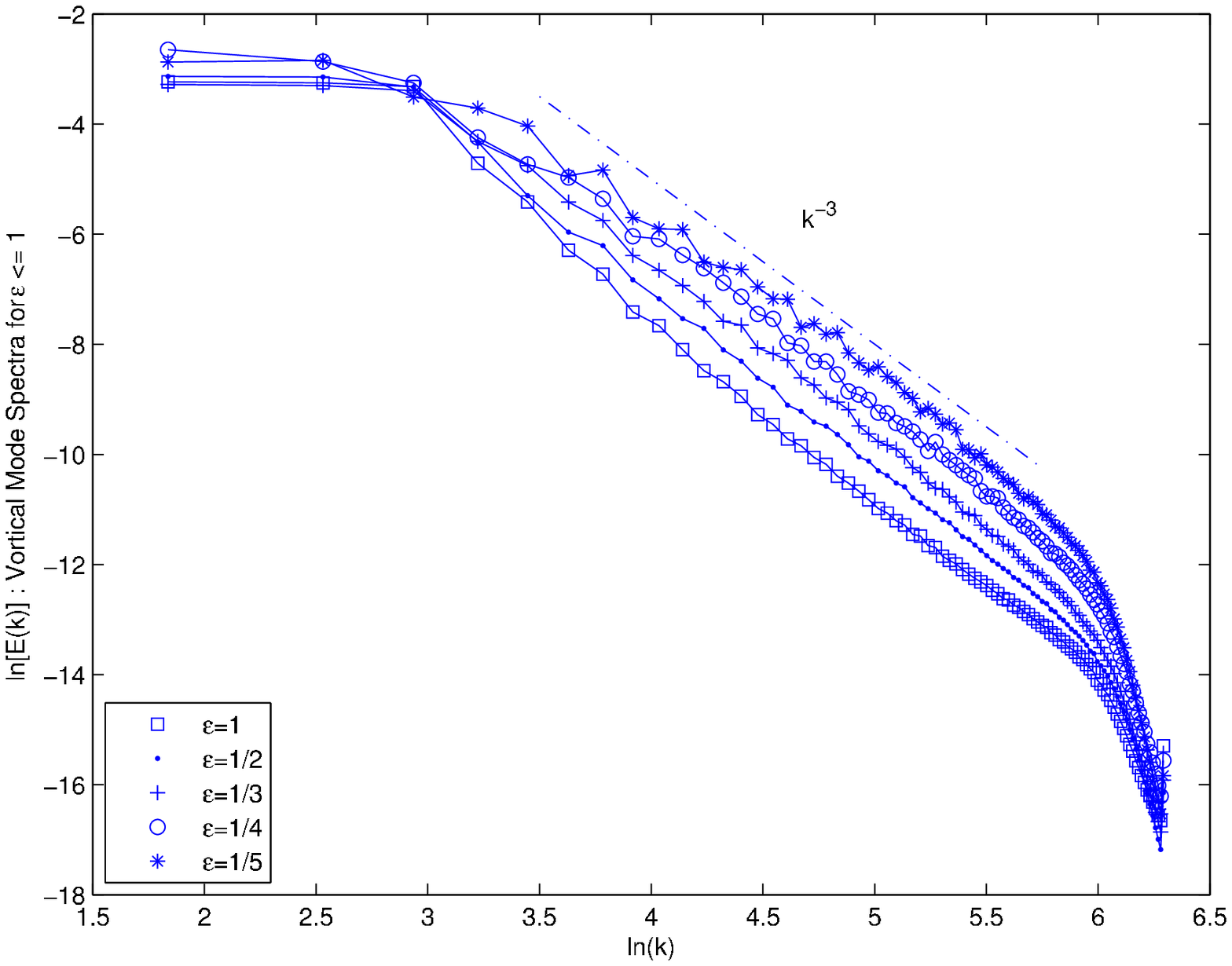}}
\centerline{\epsfxsize=10cm \epsfysize=10cm \epsfbox{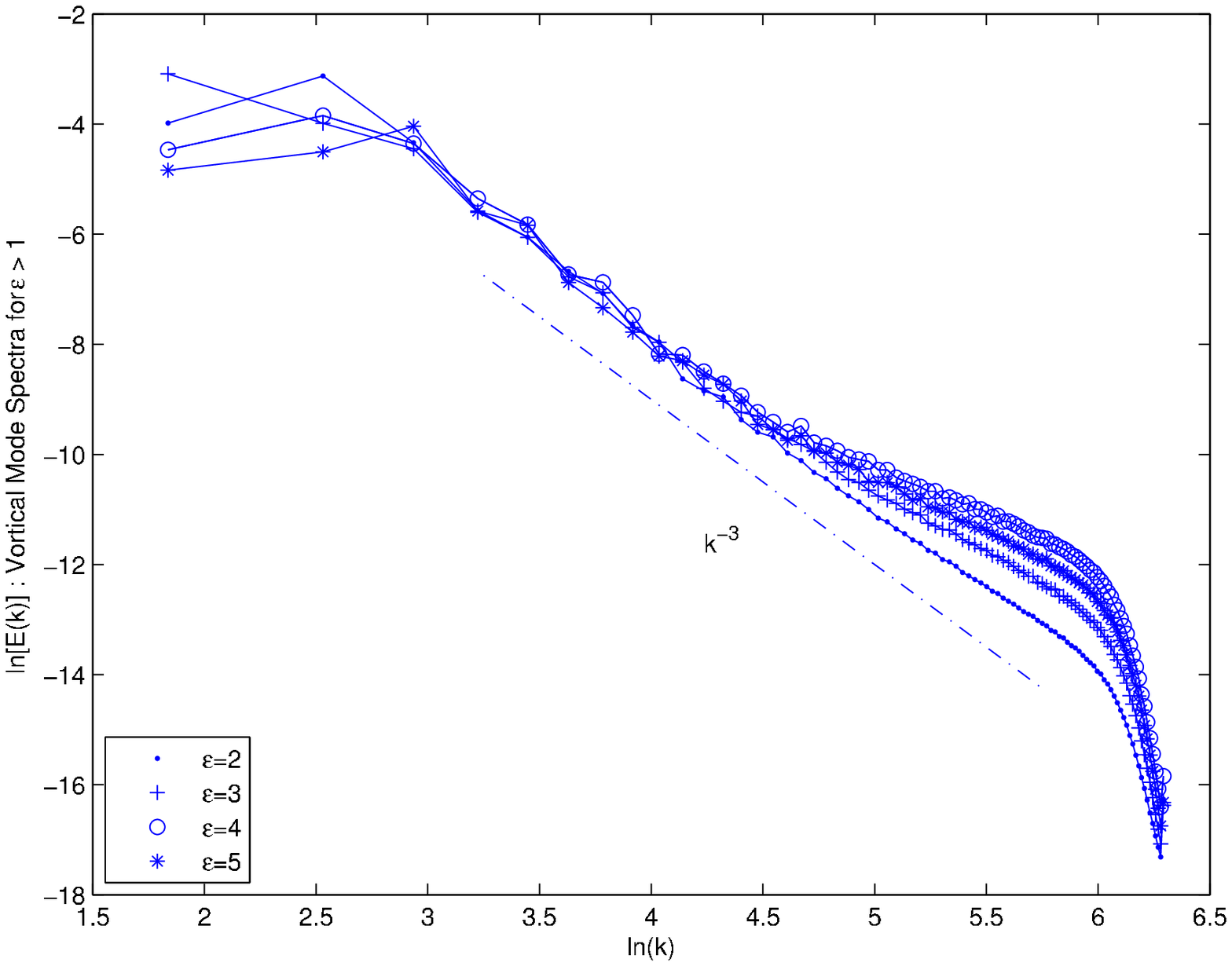}}
\caption{
Vortical mode spectra. 
The first panel shows the spectra for all $\epsilon \le 1$:  in accord with QG dynamics,
the scaling follows a $k^{-3}$ form for $k_f < k < k_d$.
The second panel shows the scaling for $\epsilon > 1$: the spectra 
again follow a $k^{-3}$ form but appear to shallow out for larger wavenumbers as $\epsilon$ increases. 
Note that
now the vortical modes contain a much smaller amount of energy than the wave modes (see Fig. (\ref{fig2})).}
\label{fig5}
\end{figure}

From Fig. (\ref{fig1}) and the first panel of Fig. (\ref{fig2}) it is evident 
that the increase in total energy at larger times for $\epsilon \le 1$ is due to the growth in the vortical modes.  As was noted above,
the wave mode energy in these cases saturates. Further,
the growth of energy at long times for $\epsilon \le 1$ is at wavenumbers 
$k< k_f$ and is the result of an inverse transfer of energy. 
As mentioned above, the simulations are 
terminated when the energy in wavenumbers 
$k< k_f$ is comparable to the energy in the forcing scale to avoid
finite size effects.
With regard to the scaling of the vortical mode spectra, 
for $\epsilon \le 1$ (first panel of Fig. (\ref{fig5}))
we observe a scaling of the form $k^{-3}$ for $k_f < k < k_d$, 
consistent with the expectations from 3D QG dynamics (Charney 1971).\\

Moving to $\epsilon > 1$, it is evident 
from the second panel of Fig. (\ref{fig2}) that the vortical modes contain a small fraction of the total energy in the system. 
In fact, the fraction of energy in the vortical modes decreases with increasing $\epsilon$,
 i.e.\ in an energetic sense they play
a progressively smaller role in the dynamics. 
Again the energy in the vortical modes increases at all scales $k< k_f$ and we stop the
simulations before finite size effects are expected to contribute.  Considering the spectra 
shown in the second panel of Fig (\ref{fig5}), the scaling
$k^{-3}$ for $k_f < k < k_2$ suggests 3D QG dynamics (though we notice the appearance 
of a shallower form at higher wavenumbers $k_2 < k < k_d$ as 
$\epsilon$ increases). 

\subsection{Total energy} 

\begin{figure}
\centering
\centerline{\epsfxsize=8cm \epsfysize=8cm \epsfbox{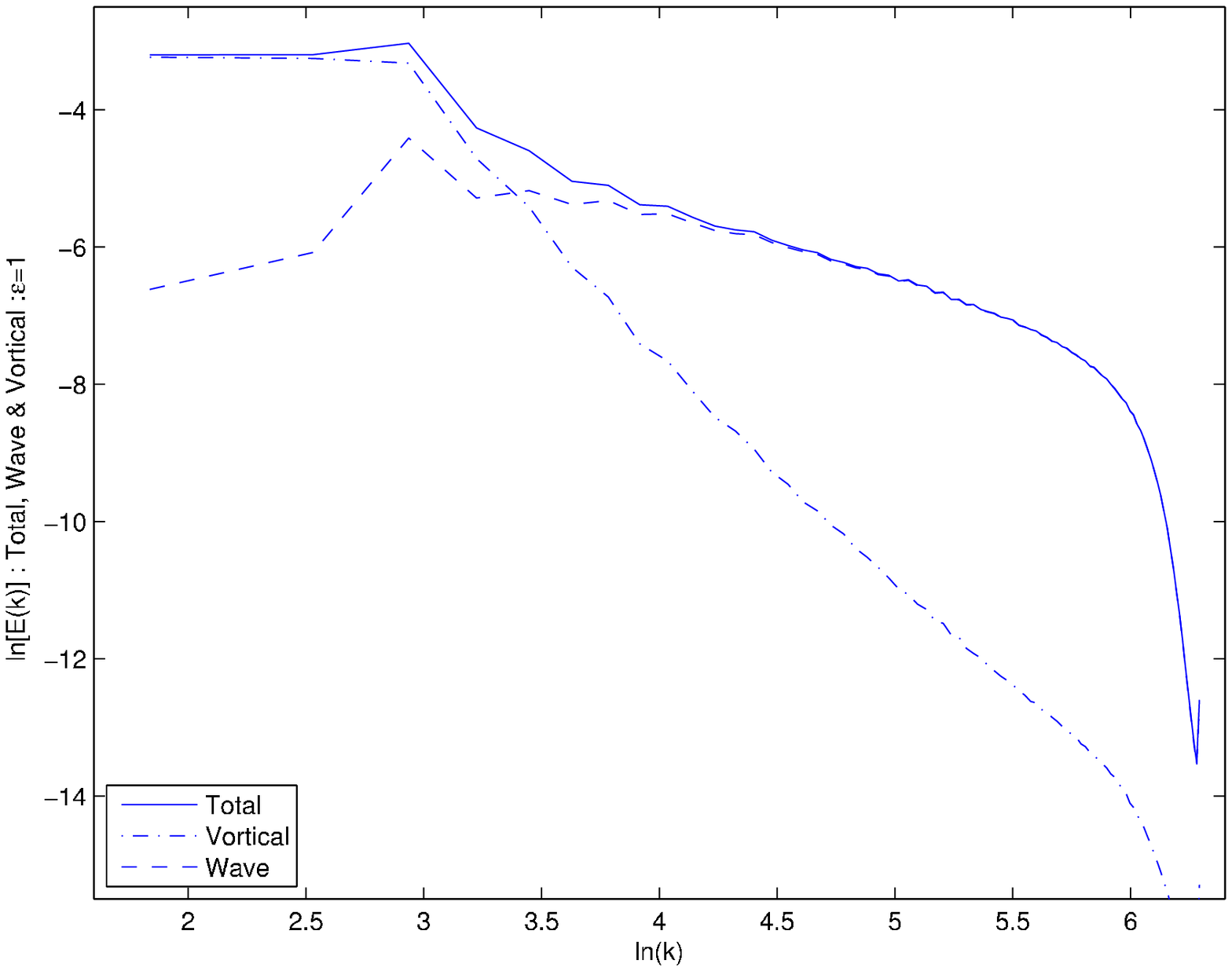}}
\centerline{\epsfxsize=8cm \epsfysize=8cm \epsfbox{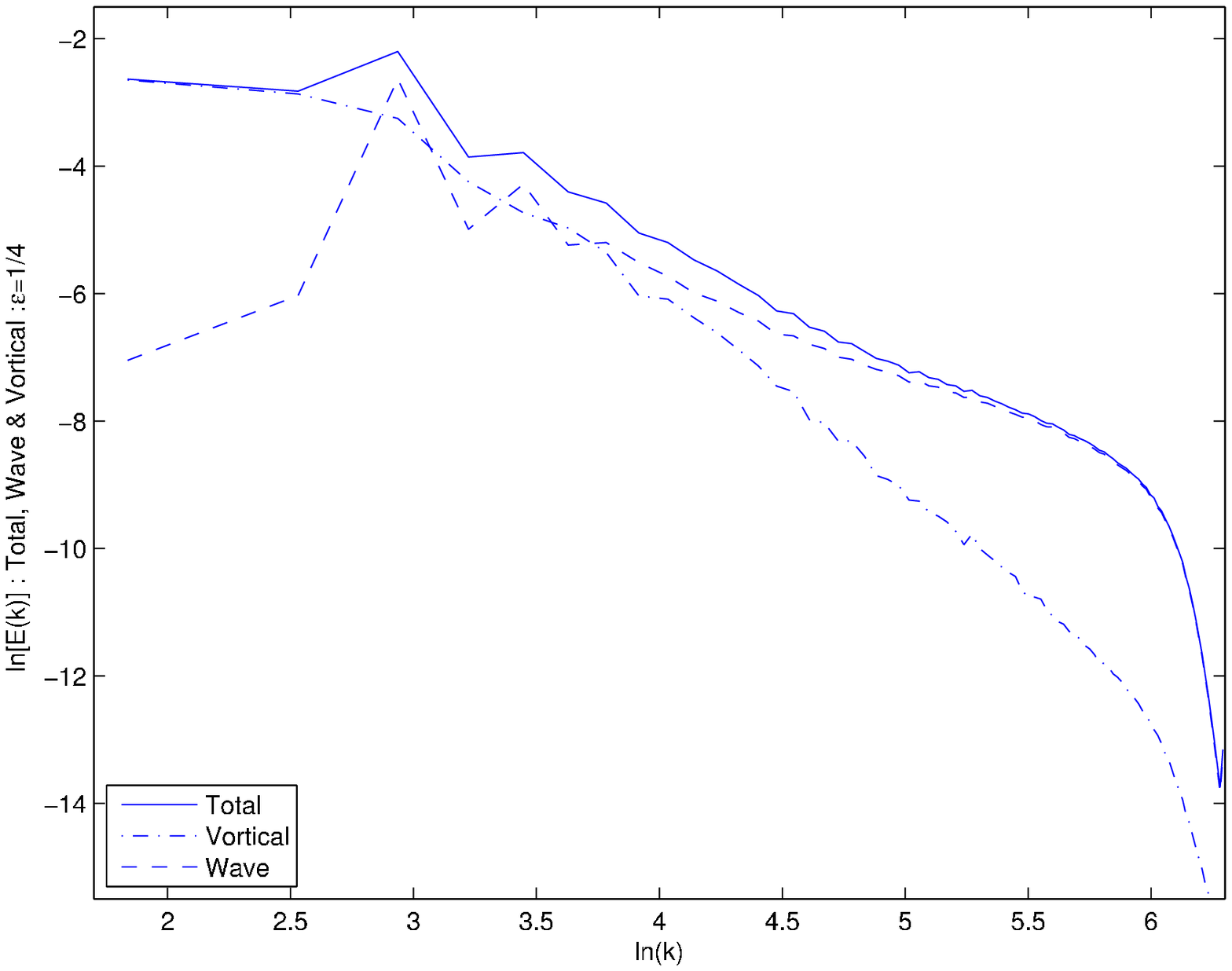}}
\centerline{\epsfxsize=8cm \epsfysize=8cm \epsfbox{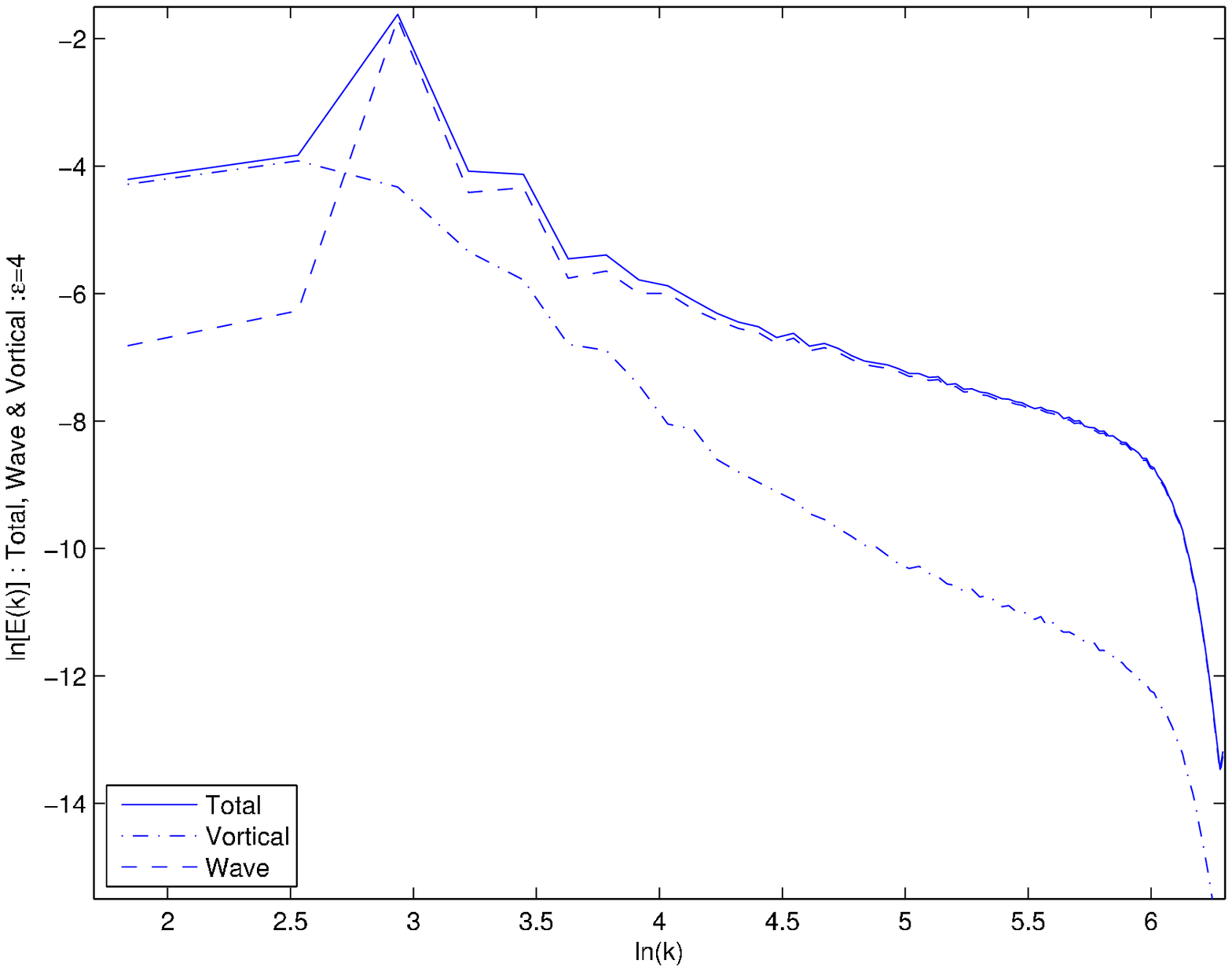}}
\caption{
Total, wave and vortical mode spectra for $\epsilon=1$ (upper), $\epsilon = 1/4$ (middle) and $\epsilon=4$ (lower) 
(other $\epsilon < 1$ and $\epsilon 
>1$ cases show respectively similar behavior). 
For $\epsilon \le 1$, the contribution of the vortical modes to the total energy 
is significant at scales $\sim k_f$, but
decreases as we move to 
progressively smaller scales --- indeed, this is particularly pronounced when $\epsilon=1$. 
On the other hand for $\epsilon=4$ (and all other $\epsilon > 1$ cases) the wave
modes dominate $\forall ~ k \ge k_f$.}
\label{fig6}
\end{figure}

Combining the vortical and wave modes, Fig (\ref{fig6}) shows the
total energy spectra. For $\epsilon \le 1$,  
we observe an unequal partition of energy among the wave and
vortical modes at varying scales. Specifically, the
vortical modes have a large fraction of the total energy at large scales, while the wave modes account for most of the energy
at small scales.
Given that the vortical mode spectrum scales more
steeply than the wave modes, this results in a gradual steep-shallow transition
in the total energy spectrum, as seen in the first two panels of Fig (\ref{fig6}).
On the other hand, for $\epsilon > 1$, the partition is always in the favor of the 
wave modes and the total energy spectra follow
the wave spectra for all $k > k_f$. 

\section{Results and discussion}

By a series of numerical simulations we have examined the effect of varying $\epsilon=f/N$ on
the vortical and wave modes 
of the 3D rotating Boussinesq equations. Specifically, we have focused on 
flows characterized by strong rotation and 
stratification, i.e.\ both $f$ and $N$ are large so as 
keep $Ro, Fr \ll 1$ and roughly comparable.  
In addition to the presence of the zero-frequency vortical mode, the
system supports wave modes with frequencies $\sigma_{\pm}(\vec{k}) \in [f,N]$. \\

For $\epsilon = f/N < 1$ (i.e.\ $Bu > 1$, stratification stronger than rotation), 
the wave mode energy saturates quite quickly.
Much like the well understood constant frequency ($\epsilon =1$) case (Bartello 1995),
the resulting forward cascade continues 
to act as efficient way of removing ageostrophic energy from the 
system. Simultaneously, the vortical modes exhibit a pronounced 
transfer of energy to large ($k<k_f$) scales, while the high wavenumber
energy spectra associated with the vortical modes scale as 
$k^{-3}$ for $k_f < k < k_d$. 
In essence, the picture painted in Bartello (1995) 
involving the 3D QG dominance of the vortical modes and 
rapid adjustment via a wave mode cascade is valid for $\epsilon < 1$. 
This picture also agrees with the analytical work by 
Babin et al.\ (1997) and Embid \& Majda (1998) regarding the 
splitting of of the vortical and wave modes for $Fr \sim Ro \rightarrow 0$ 
wherein the vortical modes 
follow 3D QG dynamics while the wave mode 
cascades are "unfrozen" and result in an efficient transfer of energy to small scales.
Given that the wave modes saturate, it is also worth noting that our results clearly indicate the 
energetic dominance of the 3D QG dynamics when both $Ro, Fr$ are small (though non-zero) and the Burger number is 
equal to or moderately larger than unity. 
Interestingly, we observe that restrictions on the resonant and near-resonant 
$(f,z,f)$ interactions when $\epsilon \neq 1$ appear to manifest themselves 
in a steepening of the high-$k$ wave mode spectra from a $k^{-1}$ scaling (a nonlocally dominated Batchelor regime where the 
fast modes are predominantly driven in a passive manner) to $k^{-5/3}$ scaling
(a local 3D turbulence like behavior where the fast modes play a dynamically active role).
Further, 
we notice a bias in the partition of total energy among the different modes at varying scales.
Specifically, the vortical modes contain a large portion of the total energy at scales $\sim k_f$ while the wave
modes account for most of the energy at smaller scales. 
This naturally introduces a gradual steep-shallow transition in 
the total energy spectrum. 
In a broader context, the focus on forward transfers of energy in the present work serves as a 
counterpart to Smith \& Waleffe (2002) wherein small scale random forcing was employed to 
probe the inverse transfer of energy for 
the regimes $1/2 \le \epsilon \le 2$ and $\epsilon \ll 1$, and to 
elucidate the resulting 3D QG or VSHF dominance, respectively. \\

Proceeding to 
$\epsilon > 1$ (i.e.\ $Bu < 1$, rotation stronger than stratification), we immediately detect an asymmetry 
in the behavior of the wave modes.
Indeed, now the 
wave modes never saturate and soon dominate the entire energy in the system. In spite of this, their spectra do appear 
to achieve near-invariance, i.e.\ they appear to retain their shape while shifting upwards. 
Further, for large $\epsilon$ there are signs of an emergent (steeper than $k^{-5/3}$) power law for 
an intermediate range of scales 
akin to the weak forward cascade in purely rotating turbulence
(Cambon et al.\ 1997, Yeung \& Zhou 1998). This situation is 
similar to the observations for the 
shallow water equations whereby switching on rotation was seen to inhibit 
forward transfer (Yuan \& Hamilton 1994; see also 
Farge \& Sadourny 1989 for remarks on the difficulty in achieving geostrophic adjustment in a similar scenario). 
As for $\epsilon \le 1$, the vortical modes continue to follow 3D QG dynamics though now their role is 
quite small since
they contain only a small fraction of the total energy in the system. In fact, observing that the 
fraction of energy in the vortical modes decreases with increasing $\epsilon$ provides some hope that, at 
least in an energetic sense, the $\epsilon \gg 1$ limit of the 3D rotating Boussinesq system will transition smoothly 
to a purely rotating flow. 
The scaling
of the vortical-mode spectra for $\epsilon > 1$ shows 
a $k^{-3}$ form for $k_f < k < k_2$ and is followed, for larger values of $\epsilon$, by a shallower form
for $k_2 < k < k_d$. 
Finally, in contrast to $\epsilon \le 1$, now the wave modes dominate the total energy in the system for all
$k > k_f$. \\

With regard to atmospheric phenomena, 
recent very high resolution studies of the rotating Boussinesq equations
with large scale forcing and $Fr \ll Ro$ in a skewed aspect ratio domain 
show a similar $k^{-3}$ scaling for vortical-mode spectra, 
$k^{-5/3}$ scaling for wave-mode spectra, 
 and vortical-mode energy dominance at large scales (Kitamura \& Matsuda 2006). 
Indeed, the spectral transitions in the total energy 
(also in the potential and kinetic energies) that occur quite naturally in 
the Boussinesq 
system (see also the discussion in Bartello 1995) when $\epsilon \le 1$ are reminiscent of the classic synoptic-mesoscale Nastrom-Gage
spectrum (Nastrom \& Gage 1985). 
However, it should be kept in mind that the $k^{-5/3}$ portion of the Boussinesq spectra 
arises from wave modes whereas some
observational evidence points to a vortical mode dominance even at small scales 
(Cho, Newell \& Barrick 1999).  Furthermore, as clearly put forth in the 
recent work of Tulloch \& Smith (2006), 
it is difficult to imagine a consistent theory of the midlatitude troposphere
that does not explicitly address the potentially complicated evolution of 
buoyancy (or potential temperature) on domain boundaries (see for example Held et al.\ 1995 and Sukhatme \& Pierrehumbert 2002). \\

{\it Acknowledgements : }
The authors gratefully acknowledge the support of NSF
CMG 0529596 and
the DOE Multiscale Mathematics program (DE-FG02-05ER25703). \\

\clearpage

\clearpage

\clearpage

\clearpage

\clearpage

\clearpage

\begin{thebibliography}{9}

\bibitem{Anile} A.M. Anile, J.K. Hunter, P. Pantano and G. Russo, {\em Ray methods for nonlinear waves in fluids and plasmas},
John Wiley and Sons NY, Pitman Monographs and Surveys in Pure and Applied Mathematics 57 (1993).

\bibitem{Ann} S.Y. Annenkov and V.I. Shrira, "Role of non-resonant interactions in the evolution of nonlinear random water wave fields,"
J. Fluid. Mech. {\bf 561}, 181 (2006).

\bibitem{Babin-rot} A. Babin, A. Mahalov and B. Nicolaenko, "Global splitting, integrability and regularity of 
3D Euler and Navier-Stokes equations for uniformly rotating fluids,"
European Journal of Mech. B Fluids, {\bf 15}, 291 (1996).

\bibitem{Babin-rot1} A. Babin, A. Mahalov and B. Nicolaenko, "Global regularity of 3D rotating Navier-Stokes equations 
for resonant domains,"
Indiana University Mathematics Journal, {\bf 48}, 1133 (1999).

\bibitem{Babin} A. Babin, A. Mahalov, B. Nicolaenko and Y. Zhou, "On the asymptotic regimes and the strongly stratified limit of
rotating Boussinesq equations,"
Theoretical and Computational Fluid Dynamics, {\bf 9}, 223 (1997).

\bibitem{Babin-rev} A. Babin, A. Mahalov and B. Nicolaenko, "Singular oscillating limits of stably-stratified
3D Euler and Navier-Stokes equations and ageostrophic wave fronts,"
in {\it Large-Scale Atmosphere-Ocean Dynamics I},
eds. J. Norbury and I. Roulstone, Cambridge University Press (2002).

\bibitem{Bartello} P. Bartello, "Geostrophic adjustment and inverse cascades in rotating stratified turbulence,"
J. Atmos. Sci. {\bf 52}, 4410 (1995).

\bibitem{Bart} P. Bartello, O. Metais and M. Lesieur, "Coherent structures in rotating three-dimensional turbulence,"
J. Fluid. Mech. {\bf 273}, 1 (1994).

\bibitem{Batch} G.K. Batchelor, "Small-scale variation of convected quantities like temperature
in turbulent fluids - Part I : General discussion and the case of small conductivity,"
J. Fluid Mech. {\bf 5}, 113 (1959).

\bibitem{Bellet} F. Bellet, F.S. Godeferd, J.F. Scott and C. Cambon, "Wave turbulence in rapidly 
rotating flows,"
J. Fluid. Mech. {\bf 562}, 83 (2006).

\bibitem{C-rev} C. Cambon, "Turbulence and vortex structures in rotating and stratified flows,"
European Journal of Mechanics - B Fluids, {\bf 20}, 489 (2001).

\bibitem{Cambon-rot} C. Cambon, N.N. Mansour and F.S. Godeferd, "Energy transfer in rotating turbulence,"
J. Fluid. Mech. {\bf 337}, 303 (1997).

\bibitem{CRG} C. Cambon, R. Rubinstein and F.S. Godeferd, "Advances in wave turbulence: rapidly rotating flows,"
New Journal of Physics, {\bf 6}, Art No. 73 (2004).

\bibitem{Charney} J.G. Charney, "Geostrophic turbulence,"
J. Atmos. Sci. {\bf 28}, 1087 (1971).

\bibitem{Chen} Q. Chen, S. Chen, G.L. Eyink and D.D. Holm, "Resonant interactions in rotating homogeneous
three-dimensional turbulence,"
J. Fluid. Mech. {\bf 542}, 139 (2005).

\bibitem{Cho} J.Y.N Cho, R.E. Newell and J.D. Barrick, "Horizontal wavenumber spectra of winds, temperature, and trace gases during the Pacific Exploratory Missions: 2. Gravity waves, quasi-two-dimensional turbulence, and vortical modes,"
Journal of Geophysical Research, {\bf 104}, D13, 16297 (1999).

\bibitem{Con} C. Connaughton, S. Nazarenko and A. Pushkarev, "Discreteness and quasiresonances in weak turbulence
of capillary waves,"
Phys. Rev. E {\bf 63}, 046306 (2001).

\bibitem{Embid1} P.F. Embid and A.J. Majda, "Averaging over wave gravity waves for geophysical flows with arbitrary
potential vorticity,"
Commun. in Partial Differential Equations, {\bf 21}, 619 (1996).

\bibitem{Embid} P.F. Embid and A.J. Majda, "Low Froude number limiting dynamics for stably stratified flow with small
or finite Rossby numbers,"
Geophys. Astrophys. Fluid Dynamics, {\bf 87}, 1 (1998).

\bibitem{Farge} M. Farge and R. Sadourny, "Wave-vortex dynamics in rotating shallow water,"
J. Fluid. Mech. {\bf 206}, 433 (1989).

\bibitem{Gill} A. Gill, {\em Atmosphere-Ocean Dynamics},
Academic Press, International Geophysics Series, Vol. 30 1982.

\bibitem{Godeferd} F. Godeferd and C. Cambon, "Detailed investigation of energy transfers in homogeneous stratified turbulence,"
Phys. of Fluids, {\bf 6}, 2084 (1994).

\bibitem{Greenspan} H.P. Greenspan, "On the nonlinear interaction of inertial mode," J. Fluid
Mech. {\bf 36}, 257 (1969). 

\bibitem{Held} I. Held, R.T. Pierrehumbert, S. Garner and K. Swanson,
"Surface quasi-geostrophic dynamics",
J. Fluid Mech. {\bf 282}, 1 (1995).

\bibitem{Jan} P.A.E.M Janssen,
"Nonlinear four-wave interactions and freak waves",
J. Phys. Oceanogr. {\bf 33}, 863 (2003).

\bibitem{Kar} E.A. Kartashova, "Weakly nonlinear theory of finite-size effects in resonances,"
Phys. Rev. Lett. {\bf 72}, 2013 (1994).

\bibitem{Kitamura} Y. Kitamura and Y. Matsuda, "The $k_h^{-3}$ and $k_h^{-5/3}$ energy spectra in stratified turbulence,"
Geophysical Research Letters, {\bf 33}, L05809 (2006).

\bibitem{Kr-p} R.H. Kraichnan, "Convection of a passive scalar by a quasi-uniform random straining field,"
J. Fluid Mech. {\bf 64}, 737 (1974).

\bibitem{LR} M.-P. Lelong and J. Riley, "Internal wave-vortical mode interactions in strongly stratified flows,"
J. Fluid. Mech. {\bf 232}, 1 (1991).

\bibitem{Lilly} D. Lilly, "Stratified turbulence and the mesoscale variability of the atmosphere,"
J. Atmos. Sci. {\bf 40}, 749 (1983).

\bibitem{Lindborg} E. Lindborg, "The energy cascade in a strongly stratified fluid,"
J. Fluid. Mech. {\bf 550}, 207 (2006).

\bibitem{Lvov} Y.V. Lvov, S. Nazarenko and B. Pokorni, "Discreteness and its effect on water wave turbulence,"
Physica D, {\bf 218}, 24 (2006).

\bibitem{Majda-book} A.J. Majda, {\em Introduction to PDEs and Waves for the Atmosphere and Ocean},
American Mathematical Society (2003).

\bibitem{Maj-Gr} A.J. Majda and M.J. Grote, "Model dynamics and vertical collapse in decaying strongly stratified flows,"
Phys. of Fluids, {\bf 9}, 2932 (1997).

\bibitem{Met} O. Metais, P. Bartello, E. Garnier, J.J. Riley, M. Lesieur, "Inverse cascade in stably stratified rotating 
turbulence,"
Dynamics of Atmospheres and Oceans, {\bf 23}, 193 (1996).

\bibitem{Herring} O. Metais and J.R. Herring, "Numerical experiments in forced stably stratified turbulence,"
J. Fluid. Mech. {\bf 202}, 97 (1989).

\bibitem{Nas} G.D. Nastrom and K.S. Gage, "A climatology of atmospheric wavenumber spectra of wind and temperature
observed by commercial aircraft,"
J. Atmos. Sci. {\bf 42}, 950 (1985).

\bibitem{Newell} A.C. Newell, "Rossby wave packet interactions,"
J. Fluid. Mech. {\bf 35}, 255 (1969).

\bibitem{RL} J. Riley and M.-P. Lelong, "Fluid motions in the presence of strong stable stratification,"
Ann. Rev. of Fluid Mech. {\bf 32}, 613 (2000).

\bibitem{RMW} J. Riley, R. Metcalfe and M. Weissman , "Direct numerical simulations of homogeneous turbulence in density-
stratified fluids," in {\it Nonlinear Properties of Internal Waves},
AIP Conference Proceedings, ed. B. West, 79 (1981).

\bibitem{Riley-new} J.J. Riley JJ and S.M. De Bruyn Kops, "Dynamics of turbulence strongly 
influenced by buoyancy,"
Phys. of Fluids {\bf 15}, 2047 (2003).

\bibitem{smith} K.S. Smith, "Comments on "The $k^{-3}$ and $k^{-5/3}$ energy spectrum of atmospheric turbulence : quasigeostrophic
two-level model simulations","
J. Atmos. Sci. {\bf 61}, 937 (2004).

\bibitem{Smith-short} L.M. Smith, "Numerical study of two-dimensional stratified turbulence,"
Contemporary Mathematics, {\bf 283}, 91 (2001).

\bibitem{SmithLee} L.M. Smith and Y. Lee, "On near resonances and symmetry breaking in forced rotating flows at moderate Rossby number,"
J. Fluid. Mech. {\bf 535}, 111 (2005).

\bibitem{SW1} L.M. Smith and F. Waleffe, "Transfer of energy to two-dimensional large scales in forced, rotating
three-dimensional turbulence,"
Phys. of Fluids, {\bf 11}, 1608 (1999).

\bibitem{SW} L.M. Smith and F. Waleffe, "Generation of slow large-scales in forced rotating stratified turbulence,"
J. Fluid. Mech. {\bf 451}, 145 (2002).

\bibitem{Jai-sqg} J. Sukhatme and R.T. Pierrehumbert, "Surface quasi-geostrophic turbulence : the study of an active scalar,"
Chaos, {\bf 12}, 439 (2002).

\bibitem{SSa} J. Sukhatme and L.M. Smith, "Eddies and Waves in a Family of Dispersive Dynamically 
Active Scalars,"
submitted to Phys. of Fluids, (2007). (referred to as 2007a)

\bibitem{SS} J. Sukhatme and L.M. Smith, "Self-similarity in decaying two-dimensional stably stratified adjustment,"
Phys. of Fluids, {\bf 19}, 036603 (2007).

\bibitem{TuS} R. Tulloch and K.S. Smith, "A theory for the atmospheric energy spectrum : Depth-limited temperature
anomalies at the tropopause,"
PNAS, {\bf 103(40)}, 14690 (2006).

\bibitem{Vann} J. Vanneste, "Wave radiation by balanced motion in a simple model,"
SIAM Journal on Applied Dynamical Systems, {\bf 5}, 783 (2006).

\bibitem{WB1} M.L. Waite and P. Bartello, "Stratified turbulence dominated by vortical motion,"
J. Fluid. Mech. {\bf 517}, 281 (2004).

\bibitem{WB3} M.L. Waite and P. Bartello, "The transition from geostrophic to stratified turbulence,"
J. Fluid. Mech. {\bf 568}, 89 (2006).

\bibitem{Watson} K.M. Watson and S.B. Buchsbaum, "Interaction of capillary waves with longer waves .1. General theory and 
specific applications to waves in one dimension," 
J. Fluid. Mech. {\bf 321}, 87 (1996).

\bibitem{Yeung} P.K. Yeung and Y. Zhou, "Numerical study of rotating turbulence with external forcing,"
Phys. of Fluids, {\bf 10}, 2895 (1998).

\bibitem{YH} L. Yuan and K. Hamilton, "Equilibrium dynamics in a forced-dissipative f-plane shallow-water system,"
J. Fluid. Mech. {\bf 280}, 369 (1994).

\bibitem{Zhou} Y. Zhou, "A phenomenological treatment of rotating turbulence,"
Phys. of Fluids, {\bf 7}, 2092 (1995).

\end{thebibliography}
\end{document}